\documentstyle[aps]{revtex}

\begin{document}

\title{The Fractal Geometry of Critical Systems}

\author{N.G. Antoniou, Y.F. Contoyiannis and F.K. Diakonos}

\address{Department of Physics, University of Athens, GR-15771 Athens,
Greece}

\maketitle

\begin{abstract}
We investigate the geometry of a critical
system undergoing a second order thermal phase transition. Using a
local description for the dynamics characterizing the system at the critical point
$T=T_c$, we reveal
the formation of clusters with fractal geometry, where the term cluster is used to
describe regions with a nonvanishing value of the order parameter. We show that,
treating the cluster as an open subsystem of the entire system, new
instanton-like
configurations dominate the statistical mechanics of the cluster. 
We study the dependence of the resulting fractal dimension on the embedding 
dimension and the scaling properties (isothermal critical exponent) of the system.
Taking into account the finite size effects we are able to calculate the size of the
critical cluster in terms of the total size of the system, the critical temperature and
the effective coupling of the long wavelength interaction at the critical point.
We also show that the size of the cluster has to be identified with the correlation 
length at criticality. Finally, within the framework of the mean field
approximation, we extend our local considerations to obtain a global description
of the system. 
\end{abstract}


\section{Introduction}

The understanding of the geometry of systems near a second order critical point 
is the subject of numerous recent works \cite{Kroger00}.  
Most of these works consider
dynamics in discrete space (lattice) and try to explain the formation of clusters
with fractal geometry on the embedding lattice in terms of the scaling
properties (critical exponents) of the system \cite{Stinch88,Coni89}.
In a recent work \cite{Anton98}
we have proposed a mechanism in order to understand the formation of the fractal clusters
at $T=T_c$ for systems defined in a continuous space. Based on a scale invariant
effective action describing the dynamics at the critical point we were mainly interested
in revealing how this dynamics leads to the formation of critical
clusters. A general class of saddle-points of the effective action at $T=T_c$ turns
out to dominate the configurations contributing to the partition function if
we consider the statistical mechanics of an open subsystem (cluster) of the global 
critical system.  
In the present work we present in more detail and completeness
the method used in \cite{Anton98} to obtain a consistent picture of the local geometry 
at the transition point for one-dimensional systems and then we apply our approach in 
order to describe critical systems in higher dimensions. We take into account also
finite-size effects and we discuss the possibility to use different functional
realizations for the order parameter characterizing the system at the critical point.
Based on the local description of the critical system we propose an algorithm, using
arguments within the framework of the mean field approximation, to construct the global
system and to determine its scaling properties.

The starting point in our investigation
 is the effective action of a thermal system
at the critical point $T=T_c$, specified in $d$ dimensions 
in terms of a macroscopic field $\phi$ (order parameter) as follows:
\begin{equation}
\Gamma_c[\phi]=g_1 \Lambda^{\beta} \int d^dx
[\frac{1}{2} ( \nabla_d \phi )^2 
+g_2 \Lambda^{\gamma} \vert \Lambda^{-d} \phi \vert ^{\delta +1}]
\label{eq:act1}
\end{equation}
In eq.~(\ref{eq:act1}) $g_1$,$g_2$ are dimensionless couplings,
$\phi \sim$ (volume)$^{-1}$ and
the ultraviolet cut-off $\Lambda$ of the underlying microscopic theory fixes 
the coarse graining scale $R_c \approx \Lambda^{-1}$ of the effective system.
Eq.~(\ref{eq:act1}) leads to the standard equation of state at $T=T_c$ : 
$$\frac{\delta \Gamma_c}{\delta \phi}\sim \phi ^{\delta}~~(\phi > 0)$$ 
and the index $\delta$ is identified with the isothermal critical
exponent of the system. The action (\ref{eq:act1}), beeing dimensionless,
implies: $\beta=-(d+2)$ and $\gamma=2d+2$. Introducing also the
dimensionless quantities:
$\hat{\phi}=\Lambda^{-d} \phi ~~~;~~~\hat{x_i}=\Lambda x_i$,
we rewrite the effective action (\ref{eq:act1}) as follows:
\begin{equation}
\Gamma_c[\hat{\phi}]=g_1 \int d^d \hat{x}
[\frac{1}{2}(\nabla \hat{\phi})^2 +g_2\vert \hat{\phi} \vert^{\delta +1}]
\label{eq:act2}
\end{equation}
The scalar quantity $\phi$ describes in general the density of an extensive
physical quantity characterizing the phase transition (like, for example,
magnetization density or particle density).

Let us now mention some examples of theories which belong to the class of
physical systems described through the effective action (\ref{eq:act2}).
\begin{itemize}
\item{{\bf{O(N) 3-d effective theory:}} The action, in the large $N$ limit,
for spherically symmetric order parameter in the internal $O(N)$ space,
is written as \cite{Wetter93}:
$$\Gamma_c[\phi]=\lambda^5 \Lambda ^{-5} \int d^3 \vec{x}[\frac{1}{2}(\nabla \phi)^2
+\frac{2}{3}(\frac{2 \pi \lambda^5}{N})^2 \Lambda^8(\Lambda^{-3} \phi)^6]$$
where $\lambda= \frac{\Lambda}{T_c}$. This action, 
for $\hat{\phi}= \Lambda^{-3} \phi$ and $\hat{\vec{x}}= \Lambda \vec{x}$, has the form:
$$\Gamma_c[\hat{\phi}]= 
\lambda^5 \int d^3 \hat{\vec{x}}[\frac{1}{2}(\nabla \hat{\phi})^2 
+\frac{2}{3}(\frac{2\pi \lambda^5}{N})^2 \vert \hat{\phi} \vert^6]$$
It belongs to the general class (\ref{eq:act2}) with : $g_1= \lambda^5$ , 
$g_2=\frac{2}{3}(\frac{2 \pi \lambda^5}{N})^2$, $d=3$, $\delta=5$.}

\item{{\bf{3-d Ising model:}} The effective action $\Gamma_c[\sigma]$ - which
describes effectively the QCD at the critical point $(T=T_c)$ \cite{Steph98} -
is written as:
$$\Gamma_c[\sigma]=T_c^{-1} \int d^3 \vec{x}[\frac{1}{2}(\nabla \sigma)^2 
+GT_c^4(T_c^{-1} \sigma)^{\delta +1}]$$ 
where the macroscopic field $\sigma \sim$ (length)$^{-1}$. This action, for
$\hat{\sigma}=\Lambda^{-1} \sigma $ and $\hat{\vec{x}}=\vec{x} \Lambda$, has the form:
$$\Gamma_c[\hat{\sigma}]= \lambda \int d^3 \hat{{\vec{x}}}[\frac{1}{2}
(\nabla \hat{\sigma})^2 +G \lambda^{\delta -3} \hat{\sigma}^{\delta +1}]$$
It belongs to the general class (\ref{eq:act2}) with $g_1=\lambda$, 
$g_2=G \lambda^{\delta - 3}$, $d=3$. We recall that $\lambda= \frac{\Lambda}{T_c}$.}

\end{itemize} 

Throughout this work we use the convention $\kappa_B=1$ (Boltzmann constant)
and the energy is given in inverse length units. We will also use, for 
simplicity, the notation $(\phi,x_i)$ instead of $(\hat{\phi}, \hat{x_i})$
meaning always, unless otherwise stated, dimensionless quantities. 

The paper is organized as follows:
In Section II we investigate the Statistical Mechanics of the critical system
for $d=1$ ($d$ is the embedding space dimension) described by an effective action
of the form (\ref{eq:act2}). The formation of fractal clusters is shown and the
corresponding geometrical characteristics (size, dimension) are determined.
In  Section III we extend the analysis  to higher dimensions.
In Section IV we study the dependence of the geometrical properties of
the critical clusters on the coarse graining scale of the effective theory.
We also apply our approach to critical systems with a more general functional form of the
order parameter. Taking care of the finite size effects we determine the
correlation length in terms of the size of the formed clusters.
Then using a mean field approach we construct the global system as a superposition of
individual clusters and we explore its scaling (geometric) properties. 
Finally in Section V we summarize our main results and we give a brief outlook.
Some lengthy formulas referred to in the main text are given in the Appendix.

\section{The 1-d system}

The statistical mechanics of the critical system is determined through the
partition function:
\begin{equation}
Z_1= \int {\cal{D}}[\phi] e^{- \Gamma_c[\phi]} 
\label{eq:pf1}
\end{equation}
The local description implies that the integration measure in eq.(\ref{eq:pf1}) is
over field configurations defined in an open ball $\Omega$ with radius $R$ and center
$\bar{x}$, subset of the space $V$ (which in fact can be infinite) occupied by the
entire system. We call a cluster $C$ the set consisting of points belonging to 
$\Omega$ ($C \subset \Omega$) for which the order parameter $\phi$ is greater than or
equal to a minimum value (cut-off) $\phi_{min}$. We identify then $\bar{x}$ with 
the center of the cluster $C$. Without loss of generality we can set $\bar{x}=0$.
The {\bf{local}} geometrical properties of the system are determined through 
the scaling properties of the extensive quantities characterizing the 
cluster $C$ as we vary the radius $R$. In order to illustrate our method we will first 
consider, for simplicity, the one-dimensional case. The extension however to higher
dimensions, as we will see in the next section, is straightforward.

In the one-dimensional case the effective action of the critical system
$(T=T_c)$, resulting from eq.(\ref{eq:act2}), is given by:
\begin{equation}
\Gamma_c[\phi]=g_1 \int_0^R dx[\frac{1}{2}(\frac{d \phi}{dx})^2 
+ g_2 \vert \phi \vert^{\delta +1}]
\label{eq:act4}
\end{equation}
Here we will consider models for which the condition $g_1 \gg 1$ is valid.
This requirement allows us to use the saddle-point approximation to evaluate
the path integration in eq.(\ref{eq:pf1}) by replacing it through a summation
over the saddle-points of the action (\ref{eq:act4}). As the subsystem $\Omega$
is {\bf{open}} no boundary conditions restrict the configurations which contribute to
the path integral (\ref{eq:pf1}). This point of view is essential in our
approach.

The saddle-point configurations $\phi(x)$ fulfil the 
Euler-Lagrange equation
$\frac{d^2 \phi}{dx^2}=-\frac{\partial U(\phi)}{\partial \phi}$
where $U(\phi)$ is the concave pontential $U(\phi) = - g_2 \vert \phi \vert^{\delta +1}$.
Considering this equation as the motion of a classical particle we get the first
order equation: 
\begin{equation}
E= \frac{1}{2}(\frac{d \phi}{dx})^2 -g_2 \vert \phi \vert^{\delta +1} 
\label{eq:energy}
\end{equation}
where $E$ is a conserved (during the classical motion) quantity identified
with the total energy of the moving particle. Eq.(\ref{eq:energy})
can be integrated (for details see the Appendix) to give, for $E=0$, 
instanton-like solutions of the form:
\begin{equation}
\phi(x)=A_1 \vert x-x_o \vert^{- \frac{2}{\delta -1}}~~~~~;~~~~~
A_1=[\frac{g_2}{2}(\delta -1)^2]^{- \frac{1}{\delta -1}}
\label{eq:inst1}
\end{equation}
with $x_o=\frac{\sqrt{2}}{(\delta -1)\sqrt{g_2}}
\left(\phi(0)\right)^{-\frac{2}{\delta -1}}$.
Thus, for $E=0$, the position of the singularity $x_o$ depends only on the initial
condition $\phi(0)$.
For $E \neq 0$ the solution 
has the form $\phi(x)=A_1 \vert x-x_o^\prime \vert^{- \frac{2}{\delta -1}}$
where now $x_o^ \prime =x_o^ \prime (\phi(0), E)$ (see the Appendix).
However, configurations with $E \neq 0$, 
contribute to the partition function $Z$ with a suppression factor 
$e^{-g_1 R \vert E \vert}$, suggesting that the dominant saddle points 
in the path summation (\ref{eq:pf1}) are those solutions of the equation of
motion for which $E \approx 0$. In this case eq.(\ref{eq:act4}) simplifies to: 
$$\Gamma_c[\phi]=2g_1g_2 \int_0^R dx \left(\phi(x)\right)^{\delta +1}$$ 
Only configurations with $x_o > R$ give a nonvanishing contribution to the
path integral (\ref{eq:pf1}). In fact, the partition function is dominated by
those saddle-points for which $x_o \gg R$. Since $x \in (-R,R)$ we can easily take:
$\phi(x)=const=A_1 x_o^{-\frac{2}{\delta -1}}$. It is straightforward to show that
these solutions correspond to the long wavelength modes of the field
$\phi(x)$ by taking the Fourier transform of eq.(\ref{eq:inst1}). We get
$~~f(k) \sim \frac{e^{ikx_o}}{(\delta -1)k^{\delta-3}}~~$ and the envelope
of $f(k)$ is given by $k \sim \frac{m}{x_o}$.

Using the above approximation  we have:
\begin{equation}
\Gamma_c=G_1 R x_o^{-2 \frac{\delta +1}{\delta -1}}
\label{eq:act5}
\end{equation}
with $G_1=2 g_1g_2 A_1^{\delta +1}$.
The summation over the saddle points of the action (\ref{eq:act4}) becomes, 
within this approximation, an ordinary integration over $x_o$ with measure:
${\cal{D}} \phi=d \mu(x_o) \approx x_o^{- \frac{\delta +1}{\delta -1}} dx_o$.
As stated above the singularity $x_o$ must  be located outside  the cluster 
$C$ to give a nonvanishing contribution to the partition function of the 
one-dimensional system. This condition fixes the lower limit in the integration
over $x_o$ to be $x_{o,min}=R$. To determine the upper limit of $x_o$ one has to 
go back to the definition of the cluster $C$ given previously. Without loss of
generality we can take the extensive quantity characterizing the geometry in $C$
to be the magnetization $M= \int_0^R \phi(x) dx$ fulfilling the condition
$M \geq \mu$ with $\mu=\int_0^R \phi_{min} dx= R \phi_{min}$. In the approximation
of constant configurations for the order parameter $\phi$ we obtain the upper
limit for $x_o$ as: $x_o \leq (\frac{AR}{\mu})^{\frac{\delta -1}{2}}$. 

The one-dimensional partition function in now written as:
$$Z_1= \int_R^{{(\frac{A_1R}{\mu})}^{\frac{\delta -1}{2}}} 
dx_o x_o^{- \frac{\delta +1}{\delta -1}} 
e^{-G_1 R x_o^{-2 \frac{\delta +1}{\delta -1}}}$$
Using this expression it is straightforward to calculate the mean value of the
magnetization in the cluster $C$:
\begin{equation}
< \int_0^R \phi(x) dx>= 
\frac{A_1}{Z} \int_R^{{(\frac{A_1R}{\mu})}^{\frac{\delta -1}{2}}} 
dx_o x_o^{- \frac{\delta +1}{\delta -1}} \left(\int_0^R 
dx A_1 x_o^{- \frac{2}{\delta -1}}\right) \cdot 
e^{-G_1 R x_o^{-2 \frac{\delta +1}{\delta -1}}} 
\label{eq:m1}
\end{equation}

Using (\ref{eq:m1}) we can show
analytically (see Section IV) that in the 
large $G_1$ limit $(G_1 \gg 1)$ there are three characteristic regions 
determining the behaviour of the integral in eq.(\ref{eq:m1}).
Setting $R_d=A_1^{- \frac{\delta +1}{\delta}} \mu^{\frac {\delta +1}{\delta}}
G_1^{\frac{1}{\delta}}$  and  $R_u=G_1^{\frac{\delta -1}{\delta +1}}$ we find
that:
\begin{itemize}
\item{
For the region $R_d \ll R \ll R_u $ we have:
$< \int_0^R \phi(x) dx> \sim R^{\frac{\delta}{\delta +1}}$, with 
coefficient $\alpha_1 \approx A_1 G_1^{- \frac{1}{\delta +1}}
\frac{\Gamma(\frac{2}{\delta +1})}{\Gamma(\frac{1}{\delta +1})}$,
leading to a
fractal structure of the cluster around the point $x=0$ with fractal mass
dimension \cite{Meakin88,Vicsek89}: $d_F= \frac{\delta}{\delta +1}$.}
\item{This behaviour crosses
over for $R \gg R_u$ to a different power law:\\
$< \int_0^R \phi(x) dx> \sim R^{\frac{\delta -3}{\delta -1}}$ suggesting the
presence of a fractal with mass dimension                           
$\tilde{d}_F=\frac{\delta -3}{\delta-1}$ at large scales.}
\item{The lower limit
$R_d$ defines a minimal scale of the critical
system, below which the fractality is broken.}
\end{itemize}
The fractality in the central region characterizes the critical system in the
sense that it corresponds to the scaling behaviour in the vicinity of the
local observer when $\mu \to 0$. The crossover scale $R_u$ gives a
measure of the correlation length of the finite system at $T=T_c$.
In Fig.~1a  we show the numerical results for the calculation of 
$<M>$ using the values $G_1=5 \cdot 10^8$ and $\delta=5$. 
We recognize the central region and the two scales $R_d$, $R_u$.
In Fig.~1b we plot, for the same values of $G_1$ and
$\delta$, the results for the mean magnetization if we ignore the breaking
at $R_d$ ($\mu \to 0$).  

\section{Fractal clusters for d $>$ 1}

Let us now extend our approach to higher dimensions starting from the
two-dimensional case. We write the effective action (\ref{eq:act2}) for
$d=2$:
$$\Gamma_c[\phi]=g_1 \int d^2 \vec{r} [\frac{1}{2} \vert \nabla \phi \vert^2
+ g_2 \vert \phi \vert^{\delta +1}]$$ 
and look for classical saddle-points in an open subset of ${\cal{R}}^2$.
The Euler-Lagrange equation, in this case, has the form:
$$\nabla^2 \phi=(\delta +1)g_2 \phi^\delta$$ 
and the corresponding instanton-like saddle-points are:
\begin{equation}
d=2~~~~,~~~~\phi_2(\vec{r})=
A_2 \vert \vec{r}- \vec{r}_o \vert^{- \frac{2}{\delta -1}}~~~~;~~~~
A_2=(\frac{g_2}{4}(\delta - 1)^2 (\delta +1))^{-\frac{1}{\delta -1}}
\label{eq:inst2d}
\end{equation}
We proceed in a similar way as for the one-dimensional case considering the 
partition function $Z_2= \int {\cal{D}}[\phi] e^{-\Gamma_c[\phi]}$.
In the path summation contribute, similarly to the 1-d case, saddle-points
for which $\vec{r}_o$ lies outside the cluster $C$. The main effect in the statistical
mechanics of the 2-d system is obtained through the summation over paths with
$\vert \vec{r}_o \vert \gg R$ ($R$ is, once again, the radius of $C$) which are
in fact constant configurations determined by the 2-d parameter $\vec{r}_o$.
In close analogy with the one-dimensional treatment we write the path integral in $Z_2$
as an ordinary integral over $\vec{r}_o$. In this regime (constant configurations)
the two-dimensional effective action is :
$$\Gamma_c=G_2 R^2 r_o^{-2 \frac{\delta +1}{\delta -1}}$$ 
with $G_2= \pi g_1[\frac{2A_2^2}{(\delta -1)^2} +g_2 A_2^{\delta +1}]=
2 \pi g_1 g_2 A_2^{\delta +1}\left(\frac{\delta +3}{4}\right)$.
Performing the calculation of the mean value of the magnetization $<M(R)>=
< \int d^2 \vec{r} \phi(\vec{r})>$, characterizing the two-dimensional critical
cluster and using the notation $R_d=A_2^{- \frac{\delta +1}{2 \delta}}
\mu^{\frac{\delta +1}{2 \delta}} G_2^{\frac{1}{2 \delta}}$ and $R_u=
G_2^{\frac{\delta -1}{4}}$ we find:
\begin{itemize}
\item{For $R_d \ll R \ll R_u$:
\begin{equation}
<M(R)> \sim R^{\frac{2 \delta}{\delta +1}}
\label{eq:twodim}
\end{equation}
with coefficient $\alpha_2 \approx \pi A_2 G_2^{- \frac{1}{\delta +1}}
\frac{\Gamma(\frac{2}{\delta +1})}{\Gamma(\frac{1}{\delta +1})}$. This suggests
the formation of a geometrical stucture in $C$ 
with fractal mass dimension $d_F=\frac{2 \delta}{\delta +1}$.}
\item{This behaviour
crosses over for $R \gg R_u$ to a power law 
$<M(R)> \sim R^{2 \frac{\delta -2}{\delta -1}}$ describing a local fractal with
mass dimension $\tilde{d}_F=2 \frac{\delta -2}{\delta -1}$ at large scales.}
\item{Finally, as in the 1-d case, for $R \ll R_d$ the fractality is broken and
the mass dimension coincides with the embedding dimension.}
\end{itemize}

The extension to dimensions $d \geq 3$ needs more care.
In this case we must take into account the relation between the
isothermal exponent $\delta$ and the anomalous dimension $\eta$:   
$\delta= \frac{d +2 -\eta}{d-2 +\eta}$ \cite{Stan87}. 
But let us first examine the case $\eta=0$. Repeating the procedure followed in
the 1-d and 2-d case we obtain analytically the saddle-points of the
$d-$dimensional critical effective action:
\begin{equation}
\phi=A_d (r_o^2 -r^2)^{\frac{2 -d}{2}}~~~~;~~~~
A_d=[\frac{d-2}{(2g_2)^{1/2}}]^{\frac{d-2}{2}} r_o^{\frac{d-2}{2}}
\label{eq:instd}
\end{equation}
and transforming the path summation in the partition function $Z_d$ into
an ordinary integration over $r_o$ we find:
\begin{equation}
<M(R)>=< \int_c d^dr \phi> \sim R^{1 + d/2}
\label{eq:pld0}
\end{equation}
for $R_d \ll R < \infty$. This means that for $d=3$ $R_u \to \infty$.
Namely, the crossover to the second fractal has disappeared. 
What happens now if we take $\eta \neq 0$ into account?
Consider the case $0 < \eta < 1$.
For a wide range of universality classes, including
the $O(4)$ theory where $\eta \approx 0.034$ \cite{Tetrad94}, the anomalous
dimension $\eta$ obeys this condition.
Actually for 3-d systems $\eta$ is very close to zero \cite{Ma76}.
The corresponding Euler-Lagrange
equation:
\begin{equation}
\nabla_d^2 \phi=(\delta +1)g_2 \phi^{\frac{d+2-\eta}{d-2+\eta}}
\label{eq:el3d}
\end{equation}
cannot be solved exactly. Only an 
approximate instanton-like solution can be obtained analytically:
\begin{equation}
\phi_d(r)=A_d(r_o^2 -r^2)^{\frac{2-d}{2}}~~~~;~~~~
A_d=(\frac{(d-2)r_o}{\sqrt{2g_2}})^{\frac{d-2}{2}}
(\frac{d-2}{\sqrt{2g_2}r_o})^{\frac{d \eta}{4}}
\label{eq:instdn}
\end{equation}
Details concerning this calculation are given in the Appendix.
Based on the solution (\ref{eq:instdn}), and following the process 
applied for one and two dimensions, we determine
$<M(R)>$ for spherical symmetric clusters in $d \geq 3$ dimensions. 
Using
$R_d=a^{- \frac{\delta +1}{d \delta}} \mu^{\frac{\delta +1}{d \delta}}
G_d^{\frac{1}{d \delta}}$ and $R_u=G_d^{- \frac{1}{d +q(\delta +1)}}$
with $a=(\frac{d-2}{\sqrt{2g_2}})^{\frac{d-2 +\frac{d \eta}{2}}{2}}$, 
$G_d=\frac{2 a^{\delta +1} \pi^d/2}{d \Gamma(d/2)}g_1g_2$ and 
$q= \frac{2-d-\frac{d \eta}{2}}{2}$, we obtain:
\begin{itemize}
\item{For $R_d \ll R \ll R_u$:
\begin{equation}
<M(R)> \sim R^{1 + \frac{d-\eta}{2}}
\label{eq:pldn}
\end{equation}
with coefficient
$\alpha_d \approx \frac{2 \pi^{d/2}}{d \Gamma(d/2)}
G_d^{-\frac{1}{\delta +1}} a \frac{\Gamma(\frac{2}{\delta +1})}
{\Gamma(\frac{1}{\delta +1})}$.} 
\item{For $R \gg R_u$ the power-law:
$$<M(R)> \sim R^{1 +\frac{d(2-\eta)}{4}}$$}
\item{Breaking of the fractality for $R \ll R_d$.}
\end{itemize}

Comparing eqs.(\ref{eq:pld0}) and (\ref{eq:pldn}) we find, for $\eta=0$, 
the same power law. This serves as a consistency check of the aproximation
we used. We calculated the saddle-points of eq.(\ref{eq:el3d}) numerically. 
We also have calculated, in the constant configuration regime, the mean
magnetization $<M(R)>$. The results are presented in Fig.~2.
In Fig.~2a we plot together the numerical and the approximate solution to the
Euler-Lagrange equation for $d=3$ and $\eta=0.34$
\footnote{We used this value istead of $\eta=0.034$ valid for $O(4)$ in order
to magnify the difference between the approximate and the numerical solution.}. 
The characteristic behaviour 
of $<M(R)>$ for $d=3$ is presented in Fig.~2b. Here we have used $G_3=10^2$ and
$\eta=0.34$. The breaking of the fractality (for $R \ll R_d$) is clearly
reproduced while the crossover is suppressed due to the small value of $\eta$.

Putting together our results for $1$, $2$ and $3$ dimensions and denoting by $d_F$ the
fractal dimension in the central region of the cluster $C$ while by $\tilde{d}_F$
the fractal dimension beyond the corresponding upper limit $R_u$ we have found:
\begin{equation}
d_F=\frac{d \delta}{\delta +1}~~~~~~;~~~~~~d=1,2,3,..
\label{eq:gfr1}
\end{equation}
\begin{equation}
\tilde{d}_F=d - \frac{2}{\delta -1}~~~~~~;~~~~~~d=1,2
\label{eq:gfr2}
\end{equation}
While for $d \geq 3$, we have:
\begin{equation}
d_F - \tilde{d}_F= \frac{\eta (d-2)}{4}~ + ~O(\eta^2)
\label{eq:dfr}
\end{equation}
A remarkable property of the geometry of the cluster is that $d_F > \tilde{d}_F$ for all 
dimensions, indicating a dilution of the cluster $C$ for distances greater than $R_u$
from the center of the cluster. In other words,
the size $R_u$ of the cluster gives a measure of the correlation length
in the finite system. For 3-d systems, however, the maximal size of a single
cluster ($R_u$) coincides practically with the size of the whole system
($\eta \approx 0$, $d_F \approx {\tilde{d}}_F$) and one recovers the
conventional behaviour of the correlation length $\xi$ in a second-order
phase transition ($\xi \sim$ size of the system). For critical systems
of low dimensionality ($d < 3$) the association of the correlation length
with the size of the system needs particular care and this issue will be
discussed in detail in Section IV.

\section{Extensions and finite-size effects}

In Section III we have shown the appearance of a fractal geometry for the cluster
$C$ in the central region of scales $R_d \ll R \ll R_u$. For $R < R_d$ we obtain 
the breaking of fractality and beyond $R_u$ a more dilute fractal emerges.
Therefore the limits $R_u, R_d$ determine the part of the
cluster with fractal dimension $d_F$. In the following we will investigate
how a change of the coarse graining scale $\Lambda$ affects the  fractality region.

We consider the transformation:
\begin{equation}
\Lambda^{-1}= \lambda{\Lambda^\prime}^{-1}
\label{eq:str}
\end{equation}
where the ultraviolet cut-off $\Lambda$ fixes the coarse-graining scale 
($\Lambda^{-1}$). Then eq.(\ref{eq:act2}) becomes:
\begin{equation}
\Gamma_c[\phi]=g_1 \lambda^{d+2} {\Lambda^\prime}^{-(d+2)} \int d^dx
[\frac{1}{2}( \nabla_d \phi)^2 +g_2 \lambda^{-(2d+2)} 
{\Lambda^ \prime}^{2d+2} 
\vert \lambda^d {\Lambda^ \prime}^{-d} \phi \vert^{\delta +1}]
\label{eq:tract2}
\end{equation}
Setting $\hat{\phi}^\prime= {\Lambda^\prime}^{-d} \phi$, 
$\hat{x}^\prime= \Lambda^\prime x$ eq.(\ref{eq:tract2}) simplifies to: 
$$\Gamma_c[\phi]=g_1 \lambda^{d+2} \int d^d \hat{x}^\prime
[\frac{1}{2}(\nabla_d \hat{\phi}^\prime)^2 +
g_2 \lambda^{d(\delta-2) + (d-2)} \vert \hat{\phi}^\prime \vert^{\delta +1}]$$ 
where the new constants $g_1^ \prime$, $g_2^\prime$ have the values:
\begin{eqnarray}
g_1^ \prime &=& g_1 \lambda^{d+2} \nonumber\\
g_2^ \prime &=& g_2 \lambda^{d(\delta -2) + (d-2)}
\label{eq:ncon}
\end{eqnarray}
We have seen that the dimensionless values of $R_u$, $R_d$ for the 1-d case
are:
$R_u=G_1^{\frac{\delta -1}{\delta +3}} \sim g_1^{\frac{\delta -1}{\delta +3}} 
g_2^{\frac{-2}{\delta +3}}$ and $R_d=A_1^{- \frac{\delta +1}{\delta}}
G_1^{\frac{1}{\delta}} \sim g_1^{\frac{1}{\delta}}g_2^{\frac{1}{\delta}}$.
According to eq.(\ref{eq:ncon}) the new values for the dimensionless limits are:
$R_u^ \prime = \lambda R_u$ and $R_d^ \prime= \lambda R_d$.
Using eq.(\ref{eq:str}) we find that the quantities $\Lambda^{-1} R_u$,
$\Lambda^{-1} R_d$ do not depend on the choice of the cutoff $\Lambda$:
\begin{eqnarray}
{\Lambda^ \prime}^{-1} R_u^ \prime &=& \Lambda^{-1} R_u   \nonumber\\
{\Lambda^ \prime}^{-1} R_d^ \prime &=& \Lambda^{-1} R_d
\label{eq:scin}
\end{eqnarray}
We may now extend these calculations for the case $d=2$ where 
$R_u = G_2^{\frac{\delta -1}{4}} \sim g_1^{\frac{\delta -1}{4}} 
g_2^{- \frac{1}{2}}$ and $R_d=A_2^{- \frac{\delta +1}{2 \delta}}
G_2^{\frac{1}{2 \delta}} \sim 
g_1^{\frac{1}{2 \delta}}g_2^{\frac{1}{2 \delta}}$ and we get again
eqs.(\ref{eq:scin}).
More complicated is the case $d \geq 3$. If we neglect the anomalous
dimension $\eta$ we recover eqs.(\ref{eq:scin}) but
taking $\eta$ into account we cannot find an analytic expression for the
limits $R_d$, $R_u$. Using,
however, the approximate solutions (\ref{eq:instdn}) we can prove the
validity of eqs.(\ref{eq:scin}) to a leading order in $\eta$.

Let us now consider the case when the order parameter is not directly
the scalar field $\phi(x)$ but a power:
$\phi^n(x)$, $n > 0$. The extensive
variable characterizing the critical geometry is now taken to be: 
$$M(R)=\int_C \phi^n(\vec{x})~d^dx$$
Performing the calculation of $<M(R)>$ at the level of the saddle-point 
approximation we obtain:
\begin{itemize}
\item{In the central fractality region 
$R_d^{(d)} \ll R \ll R_u^{(d)}$ the dimension is:
\begin{equation}
d^{(d)}_F=d (1- \frac{n}{\delta +1})
\label{eq:dfm1}
\end{equation}
where the embedding dimension $d$ takes the values: $d=1,2,3,..$.}
\item{The geometry in the external region $R \gg R^{(d)}_u$ is described
through the dimension:
\begin{equation}
\tilde{d}^{(d)}_F=d - \frac{2 n}{\delta - 1}~~~~~~;~~~~~~d=1,2
\label{eq:dfm2}
\end{equation}
valid also for $d \geq 3$ if we neglect the anomalous dimension $\eta$.}
\end{itemize}

\noindent
Using the obvious condition $\tilde{d}^{(d)}_F \geq 0$ eq.(\ref{eq:dfm2}) 
leads to an upper limit for the value of $n$:
$$n \leq \frac{d (\delta - 1)}{2}$$
Using as an example the parameter values $d=1$ and $\delta=5$
we obtain $n=1,2$ as the possible values of the power $n$. 
For this special choice of $d$ and $\delta$ we obtain a remarkable 
property when the order parameter is $\phi^2$: 
in this case ${\tilde{d}}^{(1)}_F=0$ and therefore $R^{(1)}_u$
coincides with the correlation length.
Concerning the limits $R^{(d)}_d$ and $R^{(d)}_u$ and their dependence
on the power $n$ we find:
\begin{itemize}
\item{The upper limit $R^{(d)}_u$ does not depend on $n$.}
\item{The lower limit $R^{(d)}_u$ has the following form for
a general $n$:
$$R^{(d)}_d= G_1^{\frac{n}{d (\delta + 1 - n)}} 
\left(\frac{A_d}{\mu^{1/n}}\right)^{-\frac{n (\delta + 1)}
{d (\delta + 1 - n)}} $$}
\end{itemize} 

We turn now to the question of critical cluster formation in a finite
system.
The effective action (\ref{eq:act2}) is in fact valid for the ideal
case of an infinite system. In order to take finite size effects, in a
consistent way, into account, we have to add the term $\frac{1}{2} m^2 \phi^2$ 
in (\ref{eq:act2}). In the following we will consider the statistical
mechanics, within the theoretical framework developed so far, of the modifed
effective action, which includes the above mentioned quadratic (mass) term 
in $\phi$.
Thereby we restrict ourselves in the simplified 1-d case although our
results can be extended to higher dimensions without difficulty. We also use
for the isothermal critical exponent the value $\delta=5$. The
central interest in our investigations is to determine the changes
induced to the upper limit $R_u$ of the central fractality region, due
to the presence of the mass term. This may lead us to a better understanding
of the relation between $R_u$ and the correlation length of the critical
system.

The saddle-points of the modified action are obtained (in the 1-d case) 
using the energy integral (\ref{eq:energy}). The dominant
configurations are those for which $E=0$. We have then:
\begin{equation}
\int_0^x d \xi= \pm \int_{\phi(0)}^{\phi(x)} 
\frac{d \phi}{(m^2 \phi^2 +2 g_2 \phi^6)^{1/2}}
\label{eq:eq31}
\end{equation}
Solving eq.(\ref{eq:eq31}) for $\phi$ we get:
\begin{equation}  
\phi(x)=\pm \frac{\sqrt{2 \tilde{c}} m e^{\pm m x}}{(m^2-2g_2 \tilde{c}^2
e^{\pm 4mx})^{1/2}}~~~~~~~~~;~~~~~~~~~~
\tilde{c}=\frac{\phi^2(0)}{1 + \sqrt{1+ \frac{2g_2}{m^2}\phi^4(0)}}
\label{eq:eq32}
\end{equation}
The solution (\ref{eq:eq32}) for small $m$ simplifies to:
\begin{equation}
\phi(x)=\sqrt{\frac{m}{4g_2 \tilde{c}}}\vert x-x_o \vert^{-1/2}~~~~~~~~~;
~~~~~~~~~x_o=\frac{m^2-2g_2 \tilde{c}^2}{8g_2 \tilde{c}^2 m}
\label{eq:eq34}
\end{equation}
and taking the limit $m \to 0$ we recover eq.(\ref{eq:inst1}) for $\delta=5$. 

The position of the singularity in eq.(\ref{eq:eq32}) is:
\begin{equation}
\tilde{x}_o=\pm \frac{1}{4m} ln \frac{m^2}{2 g_2 \tilde{c}^2}
\label{eq:eq36}
\end{equation}
It is easy to show that $\tilde{x}_o \to x_o$ (see eq.(\ref{eq:inst1})) 
for $m \to 0$. In the following and up to the end of this section we
will, for simplicity, drop the tilde over $x_o$. Whenever $x_o$ appears
in the following formulas it means the expression (\ref{eq:eq36}). 

Inserting eq.(\ref{eq:eq36}) in the solution (\ref{eq:eq32}) we
 finally obtain:
\begin{equation}
\phi(x)=(\sqrt{\frac{2}{g_2}}m)^{1/2} \frac{e^{m(x-x_o)}}
{(1-e^{4m(x-x_o)})^{1/2}}
\label{eq:eq37}
\end{equation}

Repeating the arguments of Section II we perform the path summation in the 
partition function of the finite system using the constant solutions, 
deduced from (\ref{eq:eq37}), for $x_o \gg x$:
\begin{equation}
\phi(x)=(\sqrt{\frac{2}{g_2}} m)^{1/2} \frac{e^{-m x_o}}{(1-e^{-4mx_o})^{1/2}}
\label{eq:eq38}
\end{equation}
Within this approximation the effective action of the finite system is:
\begin{equation}
\Gamma_c(R, x_o)=\tilde{G} R \frac{e^{-6mx_o}}{(1-e^{-4mx_o})^3}~~~~~~~~~;
~~~~~~~~~\tilde{G}=2^{5/2} g_1 g_2^{-1/2} m^3
\label{eq:eq39}
\end{equation}
The path summation in the partition function goes over to an ordinary integration 
over $x_o$ with measure obtained from eq.(\ref{eq:eq38}):
 \begin{eqnarray}
Z \sim \int_R^{x_{o_{max}}} d x_o &\phantom{a}&
[m^{3/2}e^{-m x_o}(1-e^{-4 m x_o})^{-1/2}
-m^{3/2} e^{-5mx_0}(1-e^{-4mx_0})^{-3/2}] \nonumber \\
&\phantom{a}& e^{-\tilde{G}R \frac{e^{-6m x_o}}{(1-e^{-4m x_o})^3}}
\label{eq:eq41}
\end{eqnarray}
In the limit $m \to 0$ eq.(\ref{eq:eq41}) becomes:
\begin{displaymath}
Z \sim \int_R^{x_{o max}} dx_o x_o^{-\frac{3}{2}} e^{-G_1 R x_o^{-3}}
\end{displaymath}
recovering the expression for the partition function $Z_1$ found
in Section II.

Setting now $\omega(z) =\frac{\tilde{G} R}{8 sinh^3(2m z)}$
we get from eq.(\ref{eq:eq41}):
\begin{equation}
Z \sim 
\int_{\omega(x_{o,max})}^{\omega(R)} dt f(t) e^{-t}
\label{eq:eq42}
\end{equation}
where $f(t)$ is given by:
\begin{eqnarray}
f(t)&=&m^{1/2}[(\frac{\tilde{G}R}{t})^{1/6}-(\sqrt{1+\frac{1}{4}
(\frac{\tilde{G}R}{t})^{2/3}}-\frac{1}{2}(\frac{\tilde{G}R}{t})^{1/3})
(\frac{\tilde{G}R}{t})^{1/2}] \nonumber \\
&\phantom{a}& \frac{(\tilde{G}R)^{-1}}
{\sqrt{1+\frac{1}{4}(\frac{\tilde{G}R}{t})^{2/3}}}(\frac{\tilde{G}R}
{t})^{1/3}
\label{eq:eq43}
\end{eqnarray}
The value of $R_u$ is then determined through the condition $\omega(R_u)=1$.
In the limit $m \to 0$ we find $R_u=2^{-7/4} g_1^{1/2} g_2^{-1/4}$
in accordance with the results obtained in Section II.
For $m \gg 1$ we obtain an analytic expression for $R_u$:
\begin{equation}
R_u=\frac{lnm}{2m}
\label{eq:eq46}
\end{equation}
For general $m$ however $R_u$ can be determined only
numerically. In Fig.~3 we present the results for the
dependence of $R_u$ on $m$ solving numerically the equation $\omega(R_u)=1$. 
We have used $\tilde{G}=1$. The quantity 
$\frac{1}{m}$ is actually the linear size of the critical
system. For a large system ($m^{-1} \gg 1$), the cluster size $R_u$ gets
saturated, becoming independent of $m$. The long range correlation, in this
case, is generated by succesive convolutions of neighbouring clusters. In
other words, the picture for the global system, emerging from our results
(Fig.~3), is a superposition of fractal clusters with finite size, which, by
coalescense, may create long-range ordering in the critical system. In order
to illustrate this effect, we have constructed, by simulation, a global
system in 2-d as a set of softly interacting clusters with prescribed
geometrical properties. In fact, based on a mean field approximation scheme,
it is straightforward to determine the distribution of $N$ clusters, $P(\vec{
R}_1,...\vec{R}_N)$, with centers located at the points $(\vec{R}_1...\vec{R}_N)$
in the area $S_g$ of the global system. For this purpose we consider the
pontential term in the effective action:

\begin{equation}
U(\bar{\phi}) = g_1g_2 \int_{S_r} d^2 \vec{r}^{\phantom{a} \prime} [
\bar{\phi}
(\vec{r}^{\phantom{a}\prime})]^{\delta+1} ~~~~~;~~~~~
\bar{\phi}= \frac{1}{S_r}< \int_{S_r} d^2 \vec{r}^{\phantom{a} \prime}
\phi (\vec{r}^{\phantom{a} \prime})>
\label{eq:eq47}
\end{equation}
where $S_r$ is the area occupied by a cluster of radius $r$ and the mean field
$\bar{\phi}(\vec{r})$ is written, according to our results in Section III, as
follows:
\begin{equation}
\bar{\phi}(\vec{r}) = \frac{\Gamma(\frac{2}{\delta+1})}{\Gamma(\frac{1}{\delta+1})}
(\pi g_1 g_2(\delta +3)/2)^{-\frac{1}{\delta+1}} r^{-\frac{2}{\delta+1}} ~~~~(R_d \leq
r \leq R_u)
\label{eq:eq48}
\end{equation}
From eqs.(\ref{eq:eq47}) and (\ref{eq:eq48}) we obtain:
\begin{equation}
U(\bar{\phi})= \frac{1}{2 \pi} \left( \frac{\Gamma(\frac{2}{\delta+1})}
{\Gamma(\frac{1}{\delta+1})}\right)^{\delta+1}
\left(\frac{4}{\delta +3} \right)
\int_0^{2 \pi} d \theta \ln \left(\frac{R(\theta)}{R_d}\right)
\label{eq:eq49}
\end{equation}
where $R(\theta)$ specifies the distortion of the area occupied by the cluster
in question, owing to a state of coalescense with neighbouring clusters.
Introducing the mean radius $\bar{R}=R(\bar{\theta})$ in the integral
(\ref{eq:eq49}) we finally obtain:
\begin{equation}
U(\bar{\phi})=\frac{1}{2}\left( \frac{\Gamma(\frac{2}{\delta+1})}
{\Gamma(\frac{1}{\delta+1})}\right)^{\delta+1}
\left(\frac{4}{\delta +3} \right)
\ln\left(\frac{\bar{S}}{S_d}\right)~~~~;~~~~S_d=\pi R_d^2
~~~~;~~~~\bar{S}=\pi \bar{R}^2
\label{eq:eq50}
\end{equation}
where $\bar{S}$ is a measure of the area occupied by a distorted (in general)
cluster. The distribution of $N$ clusters, in this picture, $P(\vec{R}_1...
\vec{R}_N) \sim \prod_{i=1}^N e^{-U_i(\bar{\phi})}$, is given by the following
area-law:
\begin{equation}
P(\vec{R}_1...\vec{R}_N) = Z_N^{-1} S_d^{N \alpha_{\delta}}( \bar{S}_1 \bar{S}_2
...\bar{S}_N)^{-\alpha_{\delta}}~~~~;~~~~\alpha_{\delta}=\frac{1}{2}
\left(\frac{\Gamma(\frac{2}{\delta+1})}
{\Gamma(\frac{1}{\delta+1})}\right)^{\delta+1}
\left(\frac{4}{\delta +3} \right)
\label{eq:eq51}
\end{equation}
\begin{displaymath}
Z_N= \frac{S_d^{N \alpha_{\delta}}}{N!} \int_{S_g} d^2 \vec{R}_1...d^2 \vec{R}_N
(\bar{S}_1 \bar{S}_2...\bar{S}_N)^{-\alpha_{\delta}}
\end{displaymath}

The smallness of the exponent $\alpha_{\delta}$ in eqs.(\ref{eq:eq51})
($ \alpha_{\delta} \approx 0.006 $ for $ \delta = 5 $) guarantees that the
interaction of clusters is very weak, leading to a random distribution
$P(\vec{R}_1...\vec{R}_N)$ in the area $S_g$  and to a Poisson behaviour of
the partition function, $Z_N \approx S_d^{N \alpha_{\delta}} \frac{S_g^N}
{N !}$. It permits us, therefore, to treat the global system as an almost ideal
gas of clusters. The parameters of the effective theory as well as the size
of the critical system, determine then the number of formed clusters, the
density within each cluster and the size of each cluster.

It is now straightforward to construct the global system as a set of softly
interacting clusters with prescribed geometrical properties.
For simplicity let us consider a system for which the order parameter is
interpreted as density of particles.
Given the linear size $R$, the isothermal critical exponent $\delta$ and the
effective couplings $g_1$, $g_2$ we calculate the size $R_u$ of a single
cluster and the corresponding number of clusters
$N_{cl}=\left(\frac{R}{R_u}\right)^2$ in the system.
The number of particles $n$ within each cluster 
is then given through eq.(\ref{eq:twodim}). The entire net of the clusters
is constructed in two steps. First we generate the positions of the
centers of the clusters treating them as uniformly distributed random
variables over a square with site $R$. Then we generate the points inside
each cluster with a distribution function specified by eq.(\ref{eq:twodim}).
If
two clusters (say the $i$-th and the $j$-th cluster) overlap then a point in
the $i$-th cluster is taken into account if the nearest center to it is the
center of the $i$-th cluster otherwise this point is neglected.
Such a construction is presented in Fig.~4. We have used the parameter
values: $R=1$, $\delta=5$, $g_1=50$ and $g_2=1$.

Although the fractal mass dimension of each
cluster is the same the resulting global set turns out to have a different
fractal dimension. In fact calculating the generalized dimensions of the
global set we observe that its geometric structure does not correspond to a
pure
monofractal set. Here we have used the method of factorial moments to perform
this analysis \cite{Bialas86}. We divide the region of the global system
into $M^2$ boxes of linear size $l$ ($M=\frac{R}{l}$).
Denoting by $n_i$ the number of points within the $i$-th box we
define the $p$-th order factorial moment of the distribution of the points
of the global set in space as:
\begin{equation}
F^{(p)}(M)=\frac{\frac{1}{M^2} \displaystyle{\sum_{i=1}^{M^2}} n_i
(n_i-1)..
(n_i-p+1)}{\left(\frac{1}{M^2} \displaystyle{\sum_{i=1}^{M^2}}n_i
\right)^p}
\label{eq:facmo}
\end{equation}
For a fixed value of $p$ the moment $F^{(p)}(M)$ possesses a power-law
dependence on $M$ (for $M \gg 1$) provided that the point-set under
consideration has a fractal structure: $F^{(p)}(M) \sim M^{s_p}$.
The exponent $s_p$ is related to the fractal dimension of the corresponding
point-set: $s_p=(p-1)(d-d_p)$. The dimensions $d_p$ are the generalized
dimensions characterizing the system $d_2$ beeing the corresponding average
fractal dimension. For a monofractal set the generalized dimension spectrum
is given as: $d_p = d_2$ for $p=3,4,..$. We have calculated the first three
factorial moments ($p=2,3,4$) as a function of $M$ for the set shown
in Fig.~4. The results are shown in a log-log plot in Fig.~5. We find the
exponents (slopes in the log-log plot): $s_2=0.65$, $s_3=1.61$ and
$s_4=2.6$
suggesting that the underlying set is a multifractal. A deeper understanding
of the dimension spectrum of the global system based on the construction
described above is a subject for a future investigation.

\section{Conclusions}
We have studied in detail the formation of critical clusters in a wide class
of systems, undergoing a thermal phase transition of second order.
We have used suitable, instanton-like, saddle-point configurations for the
local field-fluctuations, in order to saturate the path summation of the
partition function. In this treatment we were able to describe the critical
system locally and specify the geometrical properties of a single critical
cluster in terms of the parameters of the effective action, at the critical
point. Our main results are summarized as follows:
\begin{enumerate}
\item{In critical systems of low dimensionality $(d < 3)$ there exist two
characteristic scales $(R_d,R_u)$ which specify the geometry of any single
critical cluster, namely its maximal size ($R_u$) and its fractality region
$R_d \ll R \ll R_u$. The fractal dimension is $d_F=\frac{ \delta d}
{\delta +1}$, in agreement with other treatments on a lattice, and the
minimal scale $R_d$, below which fractality breaks down, is the analogue of
the lattice spacing in any treatment in discrete space.}
\item{Beyond the scale $R_u$ ($R \gg R_u$) the fractal dimension crosses
over to smaller values, ${\tilde{d}}_F < d_F$, and for a suitable choice
of the order parameter ($\phi^n(x)$ with $n=\frac{d (\delta - 1)}{2}$) it
may even vanish (${\tilde{d}}_F =0$). Therefore the scale $R_u$ can be
associated with the direct correlation length, $\xi_d \approx R_u
\Lambda^{-1}$ \cite{Stan87}, which coincides with the maximal size of a
single cluster.}
\item{For sufficiently large systems (size $\gg \Lambda^{-1}$) the direct
correlation length $(\xi_d \approx R_u \Lambda^{-1})$ is independent of the
size of the system (Fig.~3) and remains finite even in the thermodynamic
limit (infinite system).}
\item{The global system is built-up by a random distribution of critical
clusters which may overlap giving rise to a long-range total correlation
and therefore to density fluctuations at all scales. We have shown the
validity of such a mechanism in 2-d by developing a suitable algorithm
in order to construct the global system, based on the local description.
Our results show that the entire critical system develops strong density
fluctuations, of multifractal nature, in a wide range of scales, far beyond
the size of a single cluster (Figs.~4,5). A deeper understanding of this
global structure and in particular of the connection between the fractal
geometry of a single cluster and the multifractal spectrum of the entire
system, remains a challenging, open question.}
\item{In 3-d critical systems the anomalous dimension is approximately
zero ($\eta \approx 0$) and the crossover scale $R_u$ associated with
a single cluster coincides practically with the size of the global system
($d_F \approx {\tilde{d}}_F$). This observation leads to a simple picture
according to which the development of fluctuations at all scales, at the
critical point, is realized through the formation of self-similar clusters
at all sizes. The maximum cluster size ($R_u$) coincides with the size
of the global system and gives a measure of the correlation length which
becomes infinitely large, in the thermodynamic limit (infinite system).
In this case the geometry of the global system coincides with the geometry
of a single cluster and therefore it remains monofractal with the same
fractal dimension $d_F$.}
\end{enumerate}
Concluding, we have investigated the geometrical structure of critical
fluctuations, developed locally in a thermal system which is described
effectively by a scalar field. The formation of fractal clusters with
a mass dimension $d_F$, fixed by the universality class, has been revealed
and the mechanism for generating fluctuations at all scales, in the entire
system, based on the local properties, has been discussed. It is of interest
to note that the fluctuations of the global system obey a different geometry
in 3-d critical systems (monofractal) and in critical systems of low
dimensionality, $d<3$ (multifractal). The reason for this different behaviour
is due essentially to the fact that in 3-d systems, described by scalar
theories, the anomalous dimension $\eta$ turns out to be very small
($\eta \approx 0$).

\setcounter{equation}{0}
\renewcommand{\theequation}{{\bf E.} \arabic{equation}}
\section*{Appendix}
Integrating the first order differential equation (\ref{eq:energy})
we obtain:
\begin{displaymath}
\int_0^x d \xi=\pm \int_{\phi(0)}^{\phi(x)} \frac{d \phi}
{[2(E + g_2 \phi^{\delta+1})]^{1/2}}
\end{displaymath}
It follows that:
\begin{equation}
x=\pm \frac{1}{\sqrt{2E}}\left(\frac{E}{g_2}\right)^{1/a} I
\label{eq:app1}
\end{equation}
where:
\begin{eqnarray}
a &=& \delta +1 \nonumber\\
I &=& \int_{r_1^a}^{r_2^a} \frac{\psi^{\frac{1}{a}-1}}
{(1+\psi)^{1/2}} d \psi \nonumber\\
\psi &=& \frac{g_2}{E} \phi^a \nonumber\\
r_1 &=&\left(\frac{g_2}{E}\right)^{1/a} \phi(0) \nonumber\\
r_2 &=&\left(\frac{g_2}{E}\right)^{1/a} \phi(x) 
\label{eq:app2}
\end{eqnarray}
The integral $I$ can be determined analytically:
\begin{eqnarray}
I &=& \int_0^{r_2^a} \frac{\psi^{\frac{1}{a}-1}}{(1 + \psi)^{1/2}}d \psi -
\int_0^{r_1^a} \frac{\psi^{\frac{1}{a}-1}}{(1 + \psi)^{1/2}}
d \psi \nonumber\\
&=& a[r_2~_2F_1(1/2, 1/a; 1 +\frac{1}{a}, -r_2^a)-r_1~
_2F_1(1/2,1/a; 1 +\frac{1}{a}, -r_1^a)]
\label{eq:app3}
\end{eqnarray}
where $_2F_1$ is the hypergeometric function.

Inserting the formula \cite{Gradst}:
\begin{eqnarray*}
_2F_1(\alpha,\beta;\gamma;z) &=& \frac{\Gamma(\gamma)\Gamma(\beta-\alpha)}
{\Gamma(\beta)\Gamma(\gamma-\alpha)} (-1)^{\alpha} z^{-\alpha}~_2F_1(\alpha,
\alpha + 1-\gamma; \alpha + 1- \beta; \frac{1}{z}) \\
&+& \frac{\Gamma(\gamma) \Gamma(\alpha -\beta)}
{\Gamma(\alpha) \Gamma(\gamma - \beta)}(-1)^{\beta}z^{-\beta}~ 
_2F_1(\beta, \beta +1 - \gamma; \beta+1- \alpha; \frac{1}{z})
\end{eqnarray*}
in (\ref{eq:app3}) we obtain:
\begin{equation}
x= \pm \frac{c_1}{\sqrt{2 g_2}}( \phi(x)^{1- \frac{1}{a}} f_1- 
\phi(0)^{1-\frac{1}{a}}f_2)
\label{eq:app4}
\end{equation}
with:
\begin{equation}
c_1= \frac{\Gamma(1 +\frac{1}{a}) \Gamma(\frac{1}{a} - \frac{1}{2})}
{\Gamma(1/a) \Gamma(\frac{1}{2} + \frac{1}{a})}
=- \frac{2}{\delta-1}
\label{eq:app5}
\end{equation}
and
\begin{eqnarray}
f_1 &=& _2F_1(1/2,\frac{1}{2} - \frac{1}{a}, \frac{3}{2} - \frac{1}{a}, -\frac{E}{g_2}\phi(x)^{-a})  \nonumber \\
f_2 &=& _2F_1(1/2, \frac{1}{2} - \frac{1}{a} , \frac{3}{2} -\frac{1}{a}, -\frac{E}{g_2} \phi(0)^{-a})
\label{eq:app6}
\end{eqnarray}
Then (\ref{eq:app4}), for $E \to 0$, leads to the solution (\ref{eq:inst1}):
\begin{displaymath}
\phi(x)= \left[\frac{(\delta-1)^2 g_2}{2}\right]^{-\frac{1}{\delta-1}} 
\vert x_o - x \vert^{- \frac{2}{\delta-1}}
\end{displaymath}
with 
\begin{equation}
x_o=x_o(\phi(0))=\frac{2}{(\delta-1)
\sqrt{2g_2}}(\phi(0))^{\frac{1-\delta}{2}}
\label{eq:app7}
\end{equation}
From (\ref{eq:app4}) and for $E \neq 0$ we obtain that:
\begin{equation}
\phi(x)=\left[\frac{(\delta -1)^2 g_2}{2}\right]^{-\frac{1}{\delta-1}}
(x_o^\prime-x)^{-\frac{2}{\delta-1}}
\left(\frac{1}{f_1}\right)^{-\frac{2}{\delta-1}}
\label{eq:app8}
\end{equation}
where now $x_o^\prime=x_o f_2(E)$. Namely, if $E \neq 0$, $x_o$ depends
on two parameters:
\begin{equation}
x_o^\prime=x_o^\prime(\phi(0), E)
\label{eq:app9}
\end{equation}

{}

\vspace*{1.0cm}

\begin{center}
\large \bf {Figure Captions}
\end{center}

\noindent
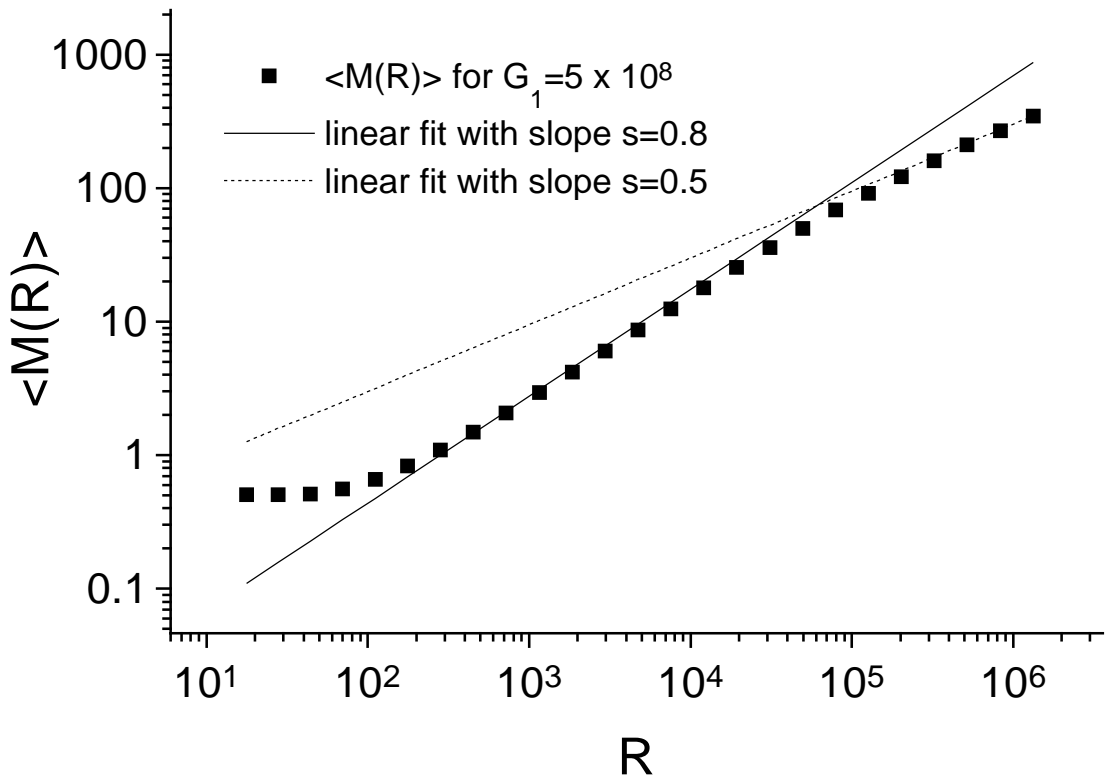
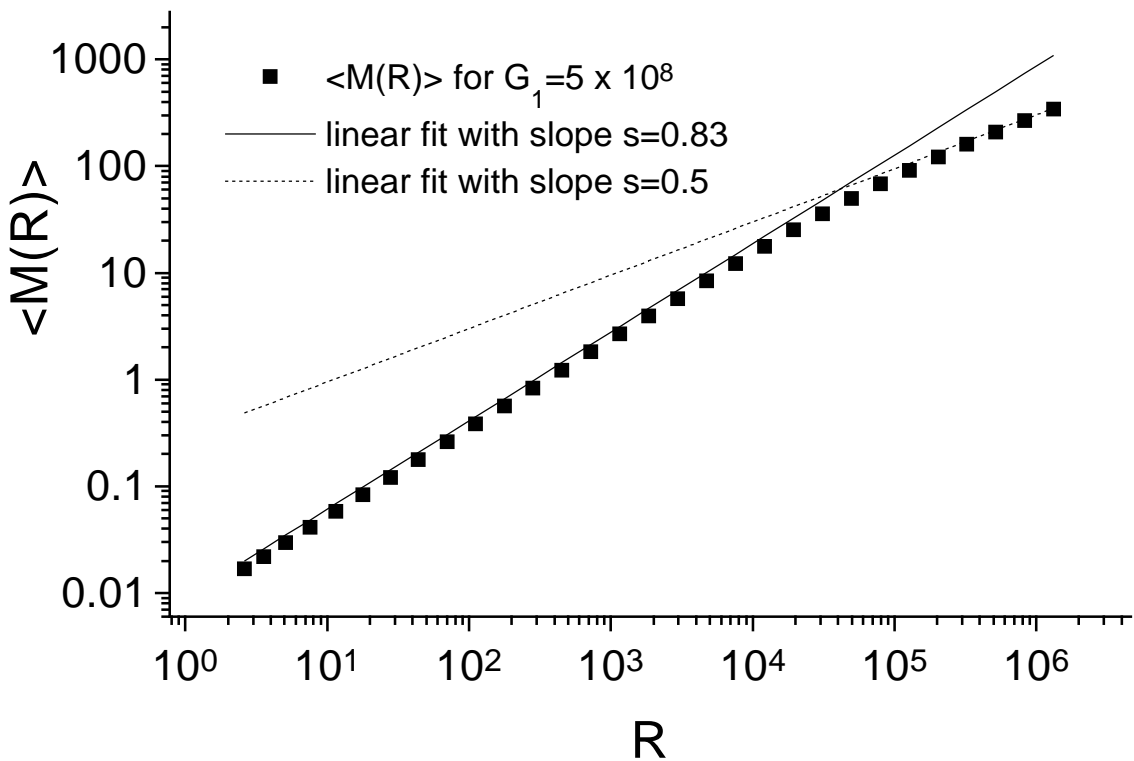
{\bf{Figure 1:}} (a) The mean magnetization $<M(R)>$ as a function of
$R$ for a 1-d critical system. The parameters are chosen so that:
$G_1=5 \cdot 10^8$ and $\mu=1$. A linear fit is also shown in order to indicate
the two different fractality regions described in Section II.  
(b) The same plot as in (a) but now with $\mu=0$. The scale $R_d$ for
the breaking of the fractality is in this case absent. All the presented
quantities are in arbitrary units.

\vspace*{0.7cm}

\noindent
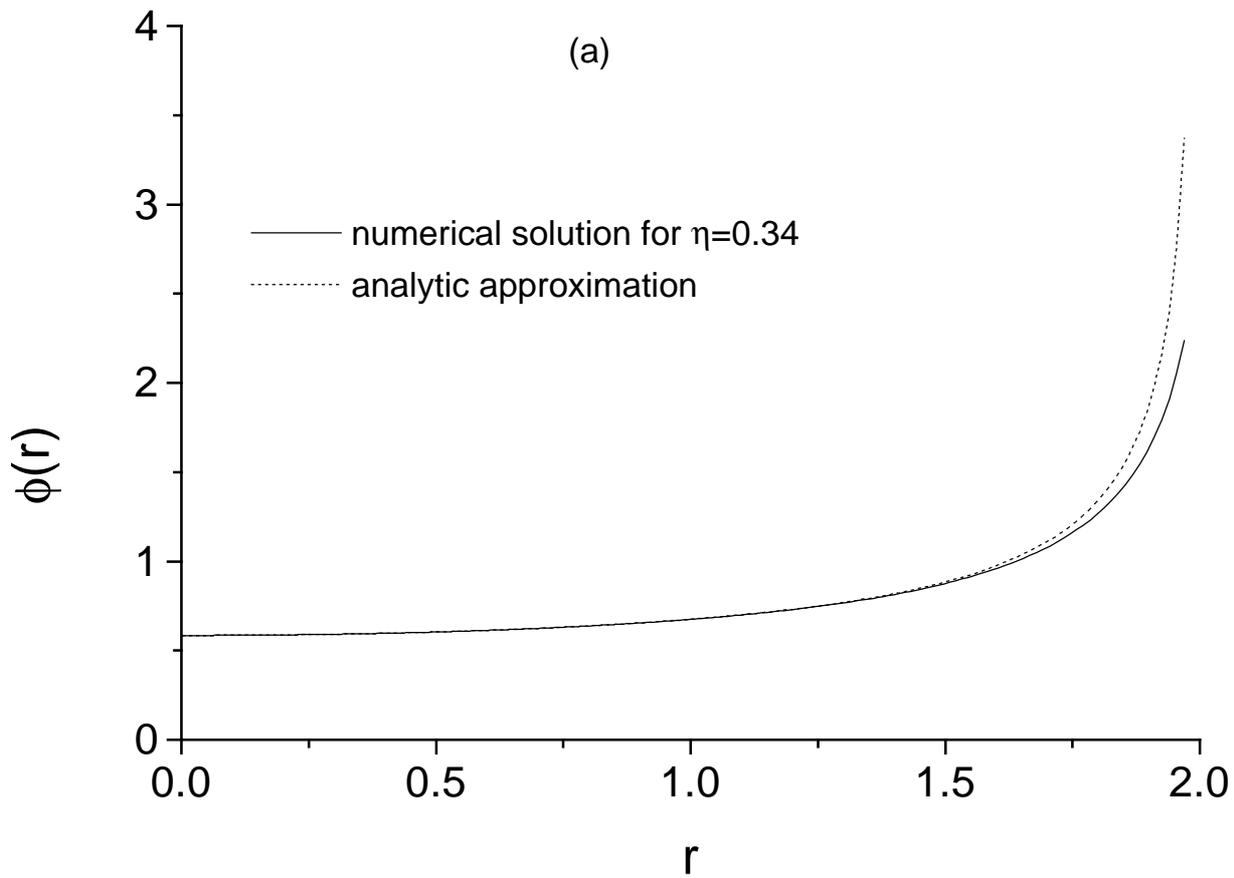
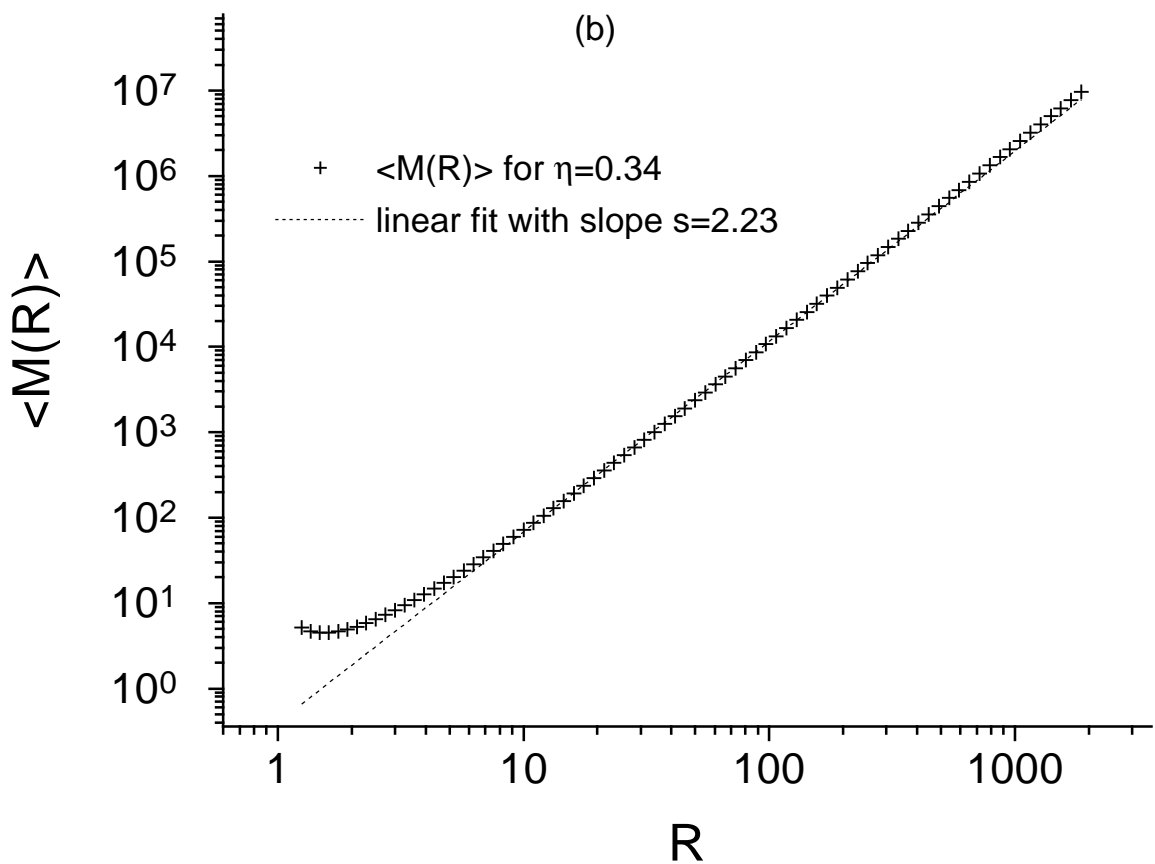
{\bf{Figure 2:}} (a) A typical 3-d instanton-like saddle-point for
$\eta=0.34$. Both the analytic approximation (dotted line) as well
as the numerical calculation (solid line) are shown.
(b) The mean magnetization $<M(R)>$ for the 3-d case using the 
saddle-points of the form presented in (a) to perform the corresponding
statistical averaging. As in Fig.~1 the displayed quantities are in arbitrary
units.

\vspace*{0.7cm}

\noindent
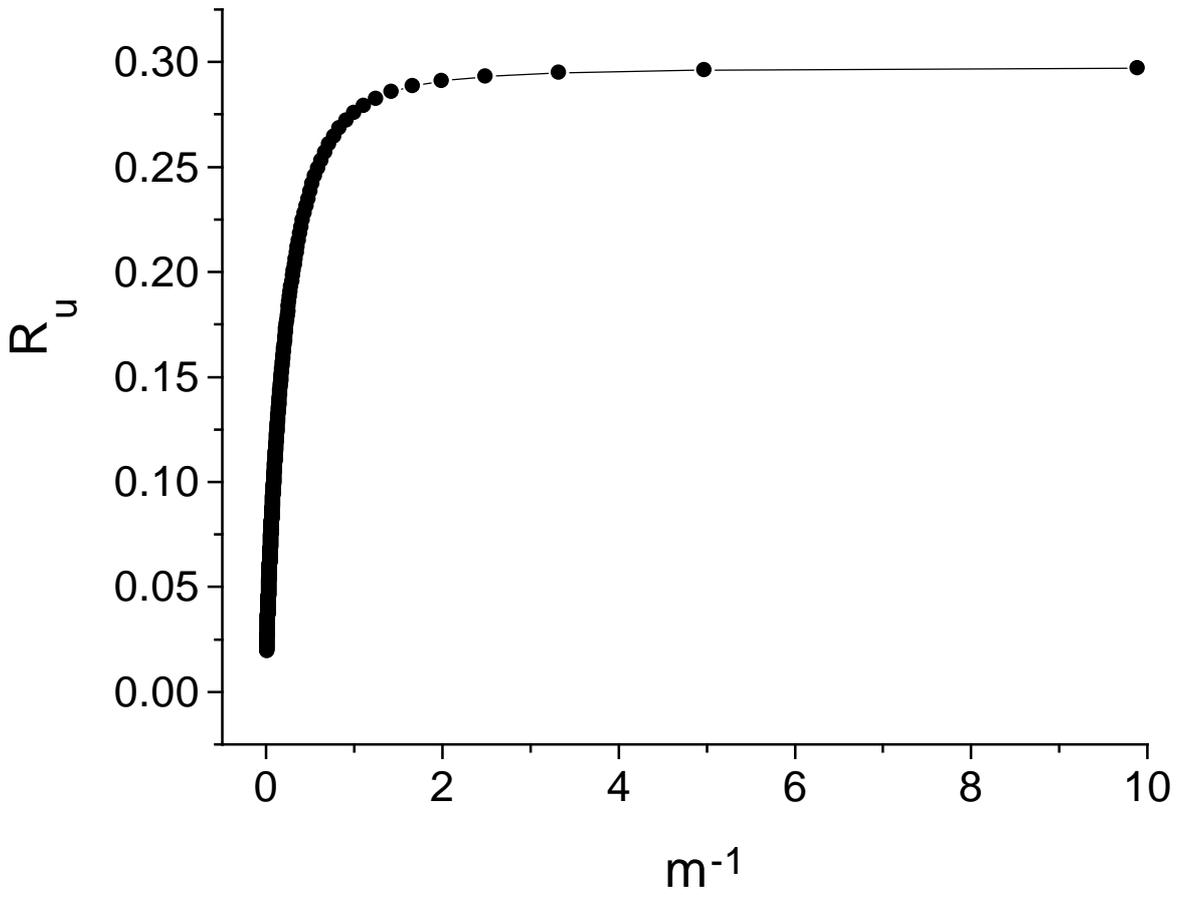
{\bf{Figure 3:}} The upper limit $R_u$ of the central fractality
region for $d=1$ as a function of the size $\frac{1}{m}$ of the critical
system. We use arbitrary units for $R_u$ and $m$.

\vspace*{0.7cm}

\noindent
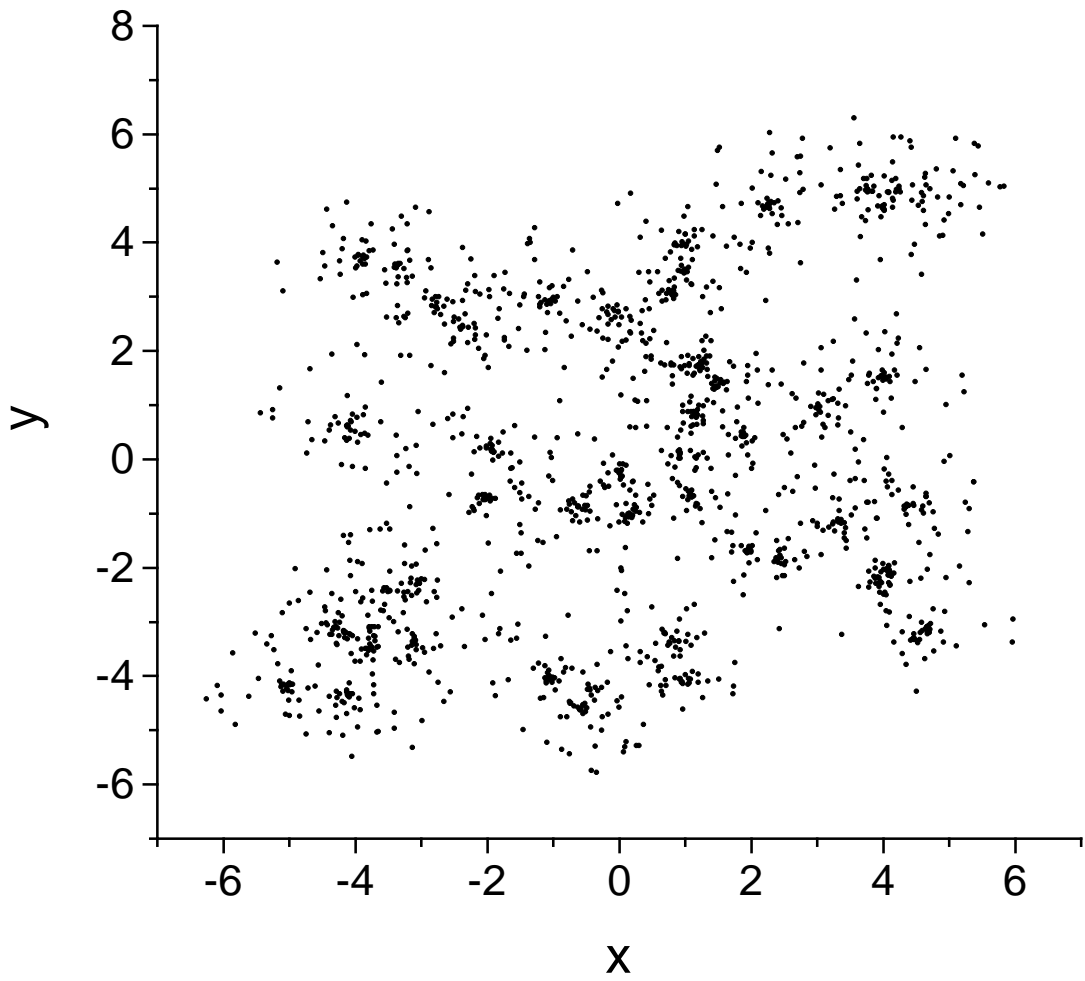
{\bf{Figure 4:}} The global 2-d critical system described through an 
effective action of the form (\ref{eq:act1}). We used the parameters:
$g_1=50$, $g_2=1$ and $\delta=5$. Each full circle represents a point
$\vec{x}$ in the 2-d space occupied by the critical system with 
$\phi(\vec{x}) > \phi_{min}$. The coordinates $x$ and $y$ are given in
arbitrary units.

\vspace*{0.7cm}

\noindent
{\bf{Figure 5:}} The first three factorial moments for the point-set
presented in Fig.~4. The linear fits indicate the slopes of the 
corresponding moments suggesting the multifractality of the
underlying set.

\end{document}